\documentclass{PoS}
\usepackage{graphicx}

\def\ap{{\sl Astropart.\ Phys.\ }}
\def\apj{{\sl Astrophys.\ J.\ }}

\def\ijmp{{\sl International\ J.\ Mod.\ Phys.\ }}

\def\mnras{{\sl MNRAS\ }}

\def\npb{{\sl Nucl.\ Phys.\ B\ }}

\def\pr{{\sl Phys.\ Rep.\ }}
\def\prd{{\sl Phys.\ Rev.\ D\ }}
\def\prl{{\sl Phys.\ Rev.\ Lett.\ }}

\def\rmp{{\sl Rev.\ Mod.\ Phys.\ }}

\newcommand{\lsim}{\,\lower2truept\hbox{${<\atop\hbox{\raise4truept\hbox{$\sim$}}}$}\,}\newcommand{\gsim}{\,\lower2truept\hbox{${>\atop\hbox{\raise4truept\hbox{$\sim$}}}$}\,}
\newcommand{\etal}{{et~al.}}

\title{The CMB as a dark energy probe}
\ShortTitle{The CMB as a dark energy probe}
\author{\speaker{Carlo Baccigalupi}\\
ITA, Albert $\ddot{\it U}$berle Strasse 2, 69120 Heidelberg, Germany, and \\
SISSA/ISAS, Via Beirut 4, 34014 Trieste, Italy\\
E-mail: \email{bacci@ita.uni-heidelberg.de, bacci@sissa.it}}
\author{Viviana Acquaviva\\
SISSA/ISAS, Via Beirut 4, 34014 Trieste, Italy\\
E-mail: \email{acqua@sissa.it}}
\FullConference{CMB and Physics of the Early Universe\\
20-22 April 2006\\
Ischia, Italy}

\abstract{We give a brief review of the known effects of a 
dynamical vacuum cosmological component, the dark energy, 
on the anisotropies of the cosmic microwave background (CMB). 
We distinguish between a ``classic" class of observables, used 
so far to constrain the average of the dark energy abundance in 
the redshift interval in which it is relevant for acceleration, 
and a ``modern" class, aiming at the measurement of its differential 
redshift behavior. \\
We show that the gravitationally lensed CMB belongs to the second 
class, as it can give a measure of the dark energy abundance at the 
time of equality with matter, occurring 
at about redshift 0.5. Indeed, the dark energy abundance at 
that epoch influences directly the lensing strength, 
which is injected at about the same time, if the source is the CMB. 
We illustrate this effect focusing on the curl (BB) component 
of CMB polarization, which is dominated by lensing on arcminute 
angular scales. An increasing dark energy abundance at the time of 
equality with matter, parameterized by a rising first order redshift derivative 
of its equation of state today, makes the BB power dropping with respect to 
a pure $\Lambda$CDM cosmology, keeping the other cosmological parameters and primordial amplitude fixed. We briefly comment on the forthcoming probes which 
might measure the lensing power on CMB.}

\begin{document}

\section{Fighting against a Cosmological Constant} 
\label{faacc}

There are conceptually two ways in which the vacuum energy manifested in the 
Einstein equations: 
\begin{equation}
G_{\mu\nu}+\Lambda g_{\mu\nu}=8\pi G T_{\mu\nu}+Vg_{\mu\nu}\ .
\label{einstein}
\end{equation} 
The first one, represented by $\Lambda$ above, has a pure geometrical conception 
and was introduced by Einstein himself for reconciling a static cosmology with 
general relativity, a mistake marked by himself as his biggest blunder. 
The second one, represented by $V$ in the equation above, was reintroduced 
afterwards by quantum mechanical arguments. \\
There is basically no expectation for $\Lambda$, while $V$ might be of the order 
of the energy scale at which all forces are predicted to unify by quantum 
gravity arguments, the Planck energy density $\rho_{Planck}$, simply because that is expected to be a fundamental scale, 
and quantum mechanics does not protect the vacuum from possessing a non-zero 
energy density. The Cosmological Constant Problem (CCP) comes from the fact that 
the vacuum energy is not compatible with the cosmological picture we guess, 
as it causes an exponential expansion which prevents structures to grow; 
this implies that the two vacuum energy terms above have to cancel out with a fantastic precision, leaving a residual which must be comparable or smaller than 
the present energy density; the latter is about 123 orders of magnitude lower than 
$\rho_{Planck}$, which implies 
\begin{equation}
\frac{|\Lambda-V|}{\rho_{Planck}}\lsim 10^{-123}\ .
\label{planck}
\end{equation} 
The CCP problem simply points out that there 
is at the present no explanation why the number above is so small, 
regardless that it is exactly zero or not. 
If one believes to the recent data from Type Ia Supernovae (SNIa,  \cite{riess_etal_1998,perlmutter_etal_1999}), Cosmic Microwave Background 
(CMB) and Large Scale Structure (LSS) combined 
(see \cite{spergel_etal_2006} and references therein), 
giving evidence for acceleration in the cosmic expansion, 
the vacuum energy density today is actually non-zero, being about 
75\% of the critical one, with a few percent precision. This means that the 
number in the equation above is non-zero, of the order of $10^{-123}$ 
with percent precision. Technically speaking, this evidence is not representing 
a new problem with respect to the CCP, but rather it is renewing the interest for it. 
The simplest explanation of the cosmic acceleration in terms of a Cosmological Constant immediately 
raises two problems, known as fine tuning and coincidence, respectively. 
Why the vacuum energy is so small with respect to the typical values in 
the early universe? Why is it comparable with the matter energy density 
at the present? In response to this embarrassment, cosmologists created 
a broader concept of cosmological vacuum component, the dark energy, 
which may be dynamic and reduces to the Cosmological Constant in the 
static limit \cite{ratra_peebles_1988,wetterich_1988}. 
The measure with the highest possible accuracy of how much the dark energy 
is close to a Cosmological Constant is likely to be one of the major challenges 
for cosmology in the forthcoming decades. In this work we make a brief review 
of the dark energy effects on the CMB anisotropies. In section \ref{pca} we 
describe how the dark energy is usually parameterized. In section 
\ref{cdaeocatritcc} we review how the CMB has been used so far to constrain 
the dark energy. In section \ref{mcmbrfdetpol} we list the new ideas of 
using the CMB as a dark energy probe, focusing on the gravitational lensing 
effect. In section \ref{lbmicmbp} we show how the lensed CMB polarization, 
and its curl component in particular, depends on the dark energy abundance 
at the onset of acceleration. Finally, in section \ref{fcmbdade} 
we briefly comment on the forthcoming CMB experimental probes. 

\begin{figure}
\centering
\includegraphics[width=12cm,height=7cm]{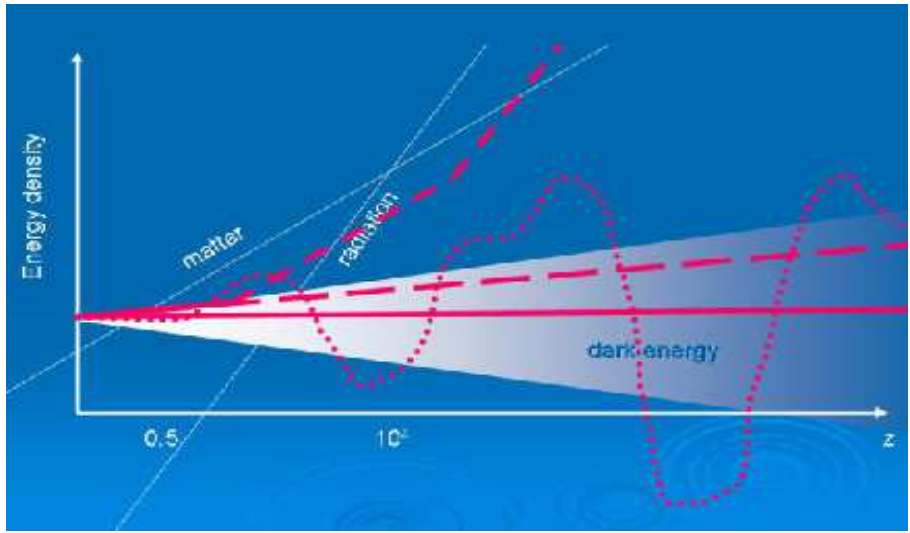}
\caption{A representation of the behavior of non-relativistic, 
relativistic matter and different models of dark energy, as 
a function of the redshift on a logarithmic scale.}
\label{abundance}
\end{figure} 

\begin{figure}
\centering
\includegraphics[width=12cm,height=7cm]{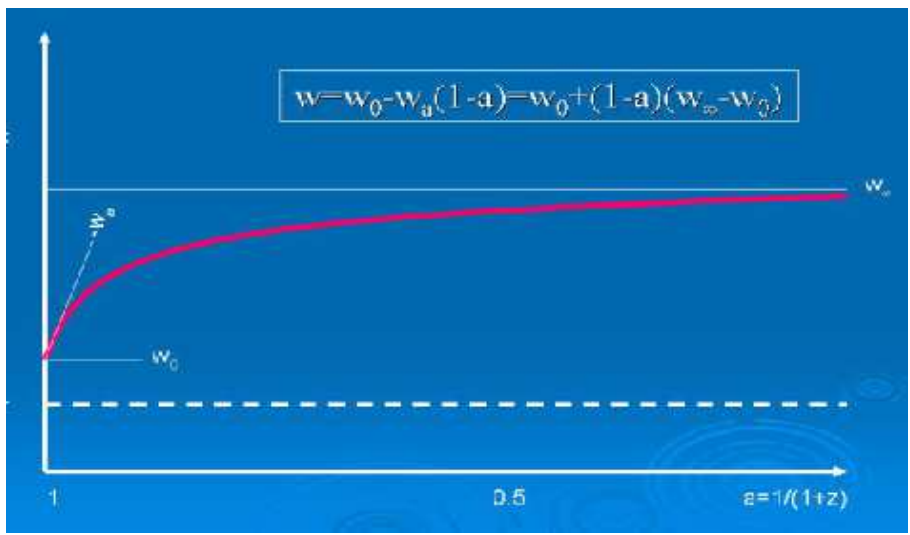}
\caption{The behavior of the dark energy equation of state 
when its value and first derivative at present are given.}
\label{figurew}
\end{figure} 

\section{Parameterizing cosmic acceleration}
\label{pca}

The main dark energy unknown is its abundance as a function of the 
redshift, $\rho(z)$. That is conveniently parameterized in terms of 
the ratio between pressure and energy density, the equation of state 
$w(z)$; the latter fixes $\rho$ by means of the Einstein conservation 
equation, which implies 
\begin{equation}
\rho=\rho_{0}\exp\left[3\int_{0}^{z}\frac{(1+w)dz'}{1+z'}\right]\ ,
\label{w}
\end{equation} 
where $\rho_{0}$ represents the present dark energy density. In figure  
\ref{abundance} we sketch the possible behaviors that the 
dark energy might have, and that have been proposed in the literature. 
The straight lines represents the scaling of non-relativistic and 
relativistic matter, as $(1+z)^{3}$ and $(1+z)^{4}$, respectively. 
The shaded area represents our ignorance on dark energy, namely the 
fact that we know with percent accuracy its abundance today 
(see \cite{spergel_etal_2006} and references therein) 
while the uncertainty increases fast with the redshift. 
Together with the Cosmological Constant behavior, the dark energy 
might follow the dominant component 
(early quintessence, short dashed line, \cite{wetterich_1988}), 
have an almost constant equation of state but larger than $-1$ 
(tracking quintessence, dashed line, \cite{ratra_peebles_1988}), or have some 
other complicated behavior, represented with dots in the figure; 
for a more complete review of the different models, see \cite{peebles_ratra_2003}. 
In the first two cases, the dark energy trajectories represent attractors 
which may solve, at least classically, the first 
of the two problems affecting the Cosmological Constant mentioned in the 
previous section. \\
If one has to measure the redshift function $\rho(z)$, 
the most unbiased approach would be to bin the redshift interval in 
which the dark energy is relevant, and provide some measure of 
$\rho$ in each bin. This approach is model independent by definition, 
has been already considered in the literature (see e.g.  \cite{crittenden_pogosian_2006,dick_etal_2006}), and will eventually 
be the standard with the increase in accuracy of the cosmological 
experiments. On the other hand, it has the disadvantage to increase the 
number of quantities to measure, with consequent decrease in accuracy 
until the data will have the appropriate quality. A viable alternative 
is represented by the parameterization of the few fundamental dynamical 
quantities that the dark energy has in any model. A convenient choice 
\cite{chevallier_polarski_2001,linder_2003} is represented by the 
present value of the equation of state, $w_{0}$ and its first derivative 
in the scale factor $w_{a}$; the latter simply measures the difference 
between $w_{0}$ and the asymptotic value in redshift, $w_{\infty}$: 
\begin{equation}
w=w_{0}-w_{a}(1-a)=w_{0}+(1-a)(w_{\infty}-w_{0})\ .
\label{modelw}
\end{equation} 
The behavior of this function is shown in figure \ref{figurew}. 
Although its appearance is markedly different from the existing 
models, represented in figure \ref{abundance}, it catches the 
relevant behavior of $w$ at least close to the present, and 
that's the reason of its wide use. 

\begin{figure}
\centering
\includegraphics[width=12cm,height=7.5cm]{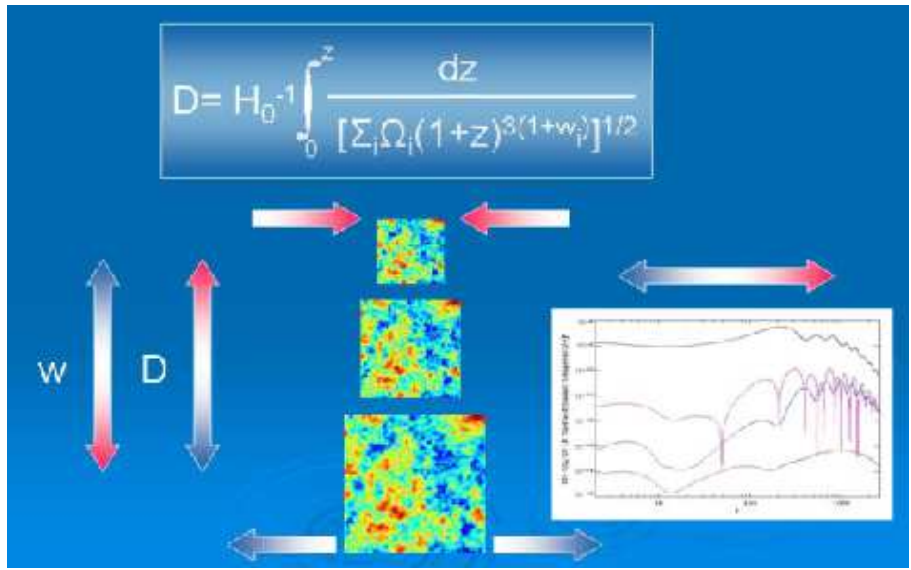}
\caption{The dark energy dynamics alter the distance to last scattering, 
causing the same features of CMB anisotropies appearing on different 
angular scales.}
\label{projection}
\end{figure} 

\begin{figure}
\centering
\includegraphics[width=12cm,height=7cm]{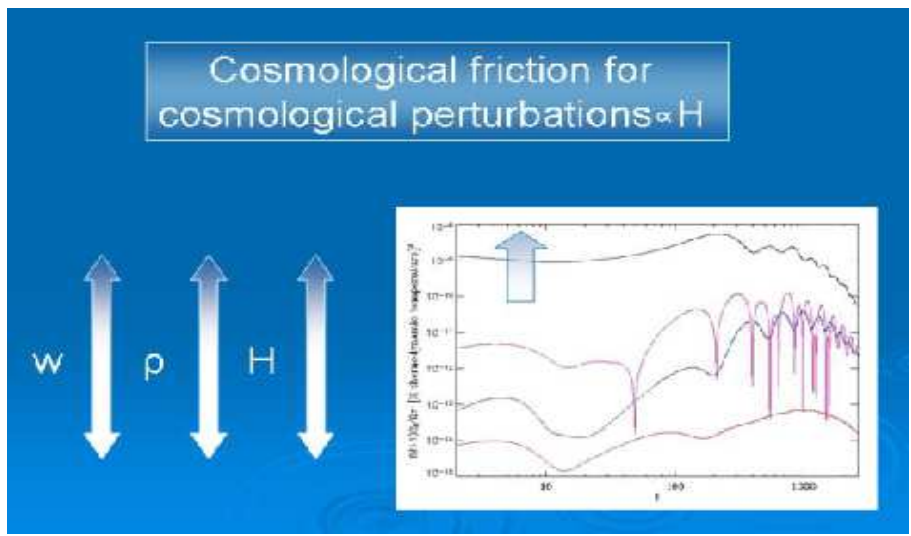}
\caption{The variation of $H$ induced by the dark energy dynamics 
boosts anisotropies on large angular scales.}
\label{isw}
\end{figure} 

\section{``Classic" dark energy effects on CMB and their role in 
the current constraints}
\label{cdaeocatritcc}

The dark energy modifies the cosmic expansion in the epoch in 
which it is relevant, i.e. at $z\lsim 0.5$. As it is easy to see 
from (\ref{w}), $w>-1$ means higher density, and consequently 
an higher Hubble parameter $H$, at all epochs starting 
from the same value today. The CMB effects 
induced by this modification were historically the first ones to be 
exploited to constrain the dark energy, and they are ``classic" in this 
sense. \\
The first one is known as projection effect, and is merely due to the 
change in the distance of the last scattering surface induced by the 
modified expansion history. The distance to the last scattering ($ls$) is 
easily evaluated as 
\begin{equation}
D=cH_{0}^{-1}\int_{0}^{z_{ls}}\frac{dz}{\sqrt{\sum_{i}\rho_{i}/\rho_{c0}}}\ ,
\label{distance}
\end{equation} 
where $c$ is the light velocity, $H_{0}$ the Hubble expansion rate today, 
$\rho_{c0}$ is the value of the cosmological critical density today, 
and the index $i$ runs over all cosmological components. Following the example 
above, $w>-1$ means that the last scattering surface gets closer, making all 
features in the CMB anisotropy appearing on larger angular scales. 
The mechanics of this effect is represented in figure \ref{projection}; 
the reported formula is a simplification of (\ref{distance}) when all 
components have a constant equation of state, and defining the parameters 
$\Omega_{i}=\rho_{i0}/\rho_{c0}$. In the bottom right panel the angular 
power spectrum of CMB anisotropies is shown for total intensity (TT, top 
curve), cross correlation between total intensity and polarization (TE, second 
curve from above), gradient and curl components of CMB polarization (EE and BB, 
third and fourth curves from above, respectively); the whole spectra shift 
to the right or left under the effect of dark energy, as the multiple $l$ 
parameterizes the inverse of the angle at which the anisotropy power is measured. \\
The second effect comes from the different behavior of cosmological 
perturbations induced by the modified expansion rate, and is known 
as Integrated Sachs-Wolfe effect (ISW). An higher $H$ 
makes an higher cosmological friction, enhancing the dynamics 
that hills and wells in the gravitational potential undergo between 
last scattering and the present. 
As photons cross them, TT CMB anisotropies are boosted, but 
only on large angular scales, since on smaller ones the effect 
is washed out by the superposition of many structures along the line of 
sight. The effect is purely metric, and does not affect polarization as 
it does not involve electromagnetic scattering of photons. It 
is illustrated in figure \ref{isw}. \\
The two effects above are both caused by the effect of the dark energy 
abundance integrated over the redshift. In the case of the projection, 
the redshift dependence of the dark energy density is actually washed 
out by the two redshift integrals (\ref{w},\ref{distance}). Clearly this 
prevents from measuring the dark energy abundance at different redshifts, 
and also introduces degeneracies with other parameters which may also 
modify the redshift averaged expansion history. Again following the 
example above, the reduction of the distance to last scattering due 
to a dark energy with $w>-1$ is also induced in closed 
cosmologies, where distances are contracted. 
Despite of these degeneracies, the CMB is a pillar of the present 
constraints on the dark energy: the combination of the data from 
LSS and CMB anisotropies represents a complementary 
probe of the expansion rate with respect to the SNIa; their combination 
allows to probe cosmologies where the dark energy equation of state is 
assumed to be a constant, indicating that in those scenarios the dark energy is a Cosmological Constant with roughly ten percent precision (see 
\cite{spergel_etal_2006} and references therein for a more accurate quotation of 
best fit confidence region): 
\begin{equation}
w=w_{0}=-1\pm 10\%\ .
\label{presentwdata}
\end{equation} 
Recently, an attempt has been made to measure the behavior of the 
dark energy at redshifts higher than one, 
exploiting the data from the Sloan Digital Sky Survey (SDSS) and 
Lyman-$\alpha$ forest; the results still indicate a Cosmological 
Constant within errors (\cite{seljak_etal_2005}, still we refer to 
the original paper for a precise quotation of errorbars): 
\begin{equation}
w_{z\simeq 0.3}=-1\pm 10\%\ \ ,\ \ w_{z\simeq 1}=-1\pm 25\%\ .
\label{pastwdata}
\end{equation} 
The effects we just described are largely used and known because 
they are common to all dark energy models. On the other hand there may 
be others arising in specific scenarios. For example, in early quintessence 
cosmologies where a non-vanishing dark energy density is present at last 
scattering, the latter process results affected, with consequent modification 
of the shape of the acoustic peaks \cite{doran_etal_2001}. 

\begin{figure}
\centering
\includegraphics[width=12cm,height=7cm]{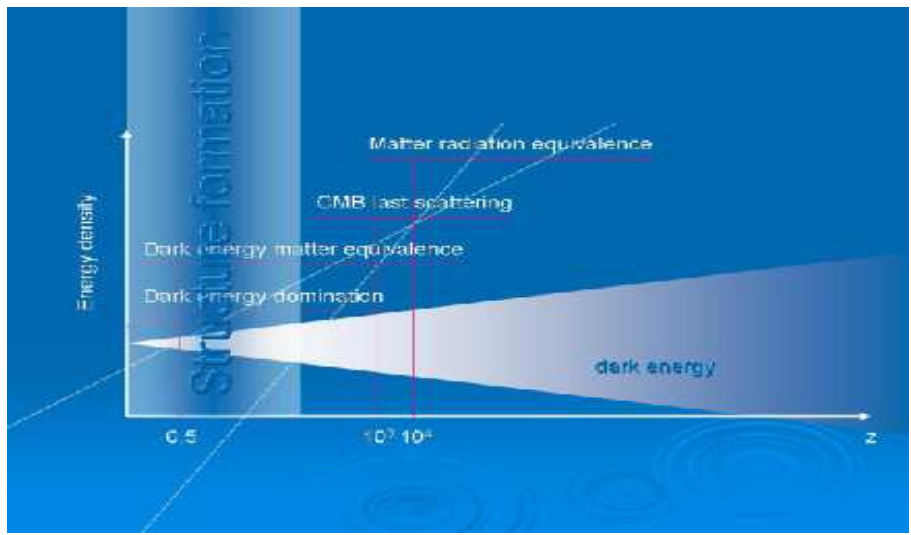}
\caption{A representation of some relevant epochs in cosmology.}
\label{epochs}
\end{figure} 

\begin{figure}
\centering
\includegraphics[width=12cm,height=7cm]{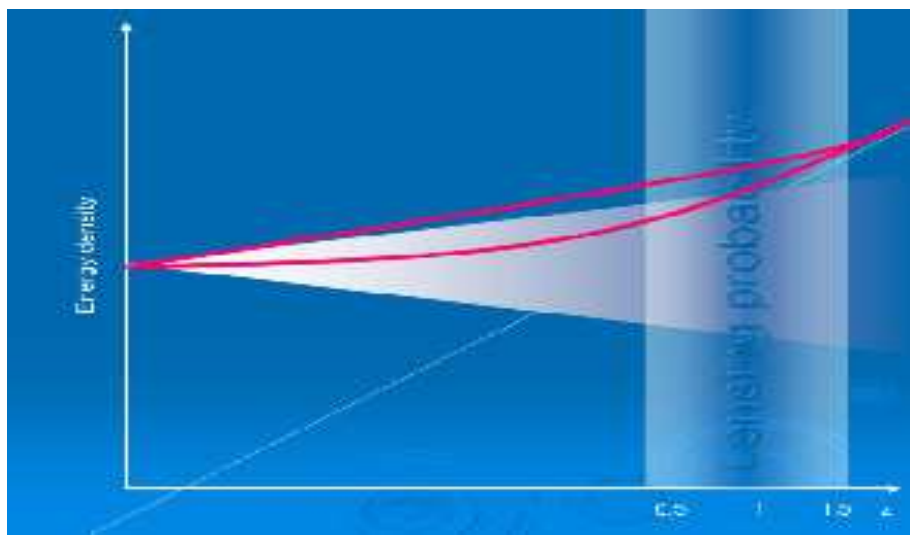}
\caption{A representation of the lensing probability, 
distinguishing dark energy models at high redshifts.}
\label{CMBlensing}
\end{figure} 

\section{``Modern" CMB relevance for dark energy: the promise of lensing}
\label{mcmbrfdetpol}

Despite of the remarkable experimental results exposed above, the theoretical 
difficulties related to the Cosmological Constant represent a motivation to 
push the battle a little forward. The entire redshift behavior of the dark 
energy has to be measured in the forthcoming years, at all epochs relevant 
for acceleration, going well beyond the present data on its redshift average. 
To this purpose, several probes are being studied in addition to the improvement 
of the data from SNIa, CMB and LSS, opening the way to the ``modern" era of dark 
energy observations. In figure \ref{epochs} we show a little scheme of the relevant  cosmological epochs. The onset of acceleration is a recent process in terms of redshift, and overlaps with structure formation. For this reason most of the modern dark energy probes try to exploit such overlap. Several authors are studying the dark energy relevance of effects like the oscillations imprinted by baryons in the dark matter distribution  \cite{white_2005}, the correlation between the ISW effect on the CMB and the observed LSS  distribution \cite{pogosian_etal_2005}, the gravitational lensing of galaxies \cite{hu_jain_2004} and CMB 
\cite{acquaviva_baccigalupi_2005}; all these works are the latest ones of 
several by many authors which may be found in the references. \\
In this paper we make some comments on the last aspect only; the CMB lensing 
relevance for dark energy comes from an elementary geometric property of the gravitational lensing process: the lensing probability is zero if the lens position coincides with the observer or the source position. Therefore, any lensing observable picks up its signal roughly in the middle between source and observer. As sketched in figure \ref{CMBlensing}, if the source is the CMB the lensing power peaks at about 
redshift 1 (see \cite{bartelmann_schneider_2001} and references therein); 
this is very interesting for dark energy, because it means that the CMB lensing is 
potentially relevant for constraining its abundance at the onset of acceleration, 
independently on its present behavior. \\
The weak lensing of the CMB by large scale cosmological structures along the line of sight is a science per se, see e.g. \cite{lewis_challinor_2006} and references therein: it is relevant on the typical angular scales subtended by cosmological structures at $z\simeq 1$, say from a few arcminutes to the degree, and represents 
a second order cosmological effect, caused by perturbations in the matter density 
onto CMB anisotropies. Therefore it correlates different scales, making the overall 
CMB statistics non-Gaussian, smearing out the acoustic peaks in the angular power 
spectrum, transferring power from the EE component of the CMB polarization 
to the BB one. This last effect is the one we focus on in the following. 

\begin{figure}
\centering
\includegraphics[width=12cm,height=7.5cm]{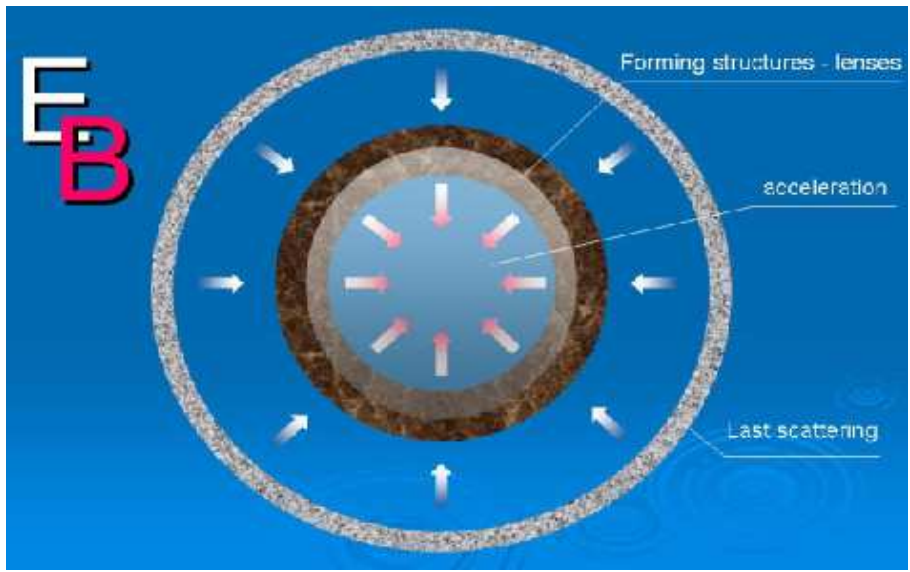}
\caption{A sketch of the generation of lensing BB modes in the CMB polarization 
anisotropies.}
\label{BB}
\end{figure} 

\section{Lensing BB modes in CMB polarization}
\label{lbmicmbp}

\begin{figure}
\centering
\includegraphics[width=7cm,height=5cm]{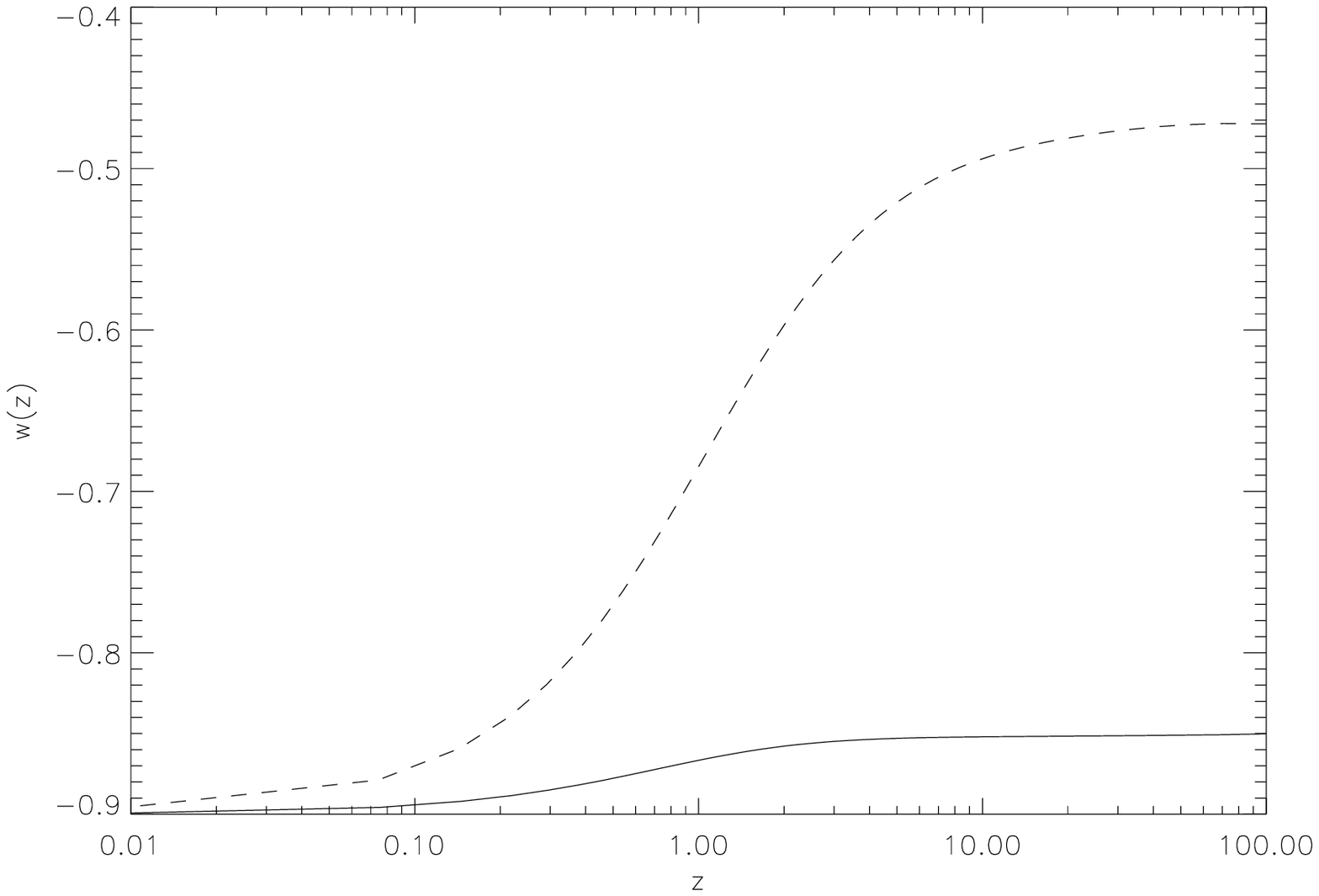}
\includegraphics[width=7cm,height=5cm]{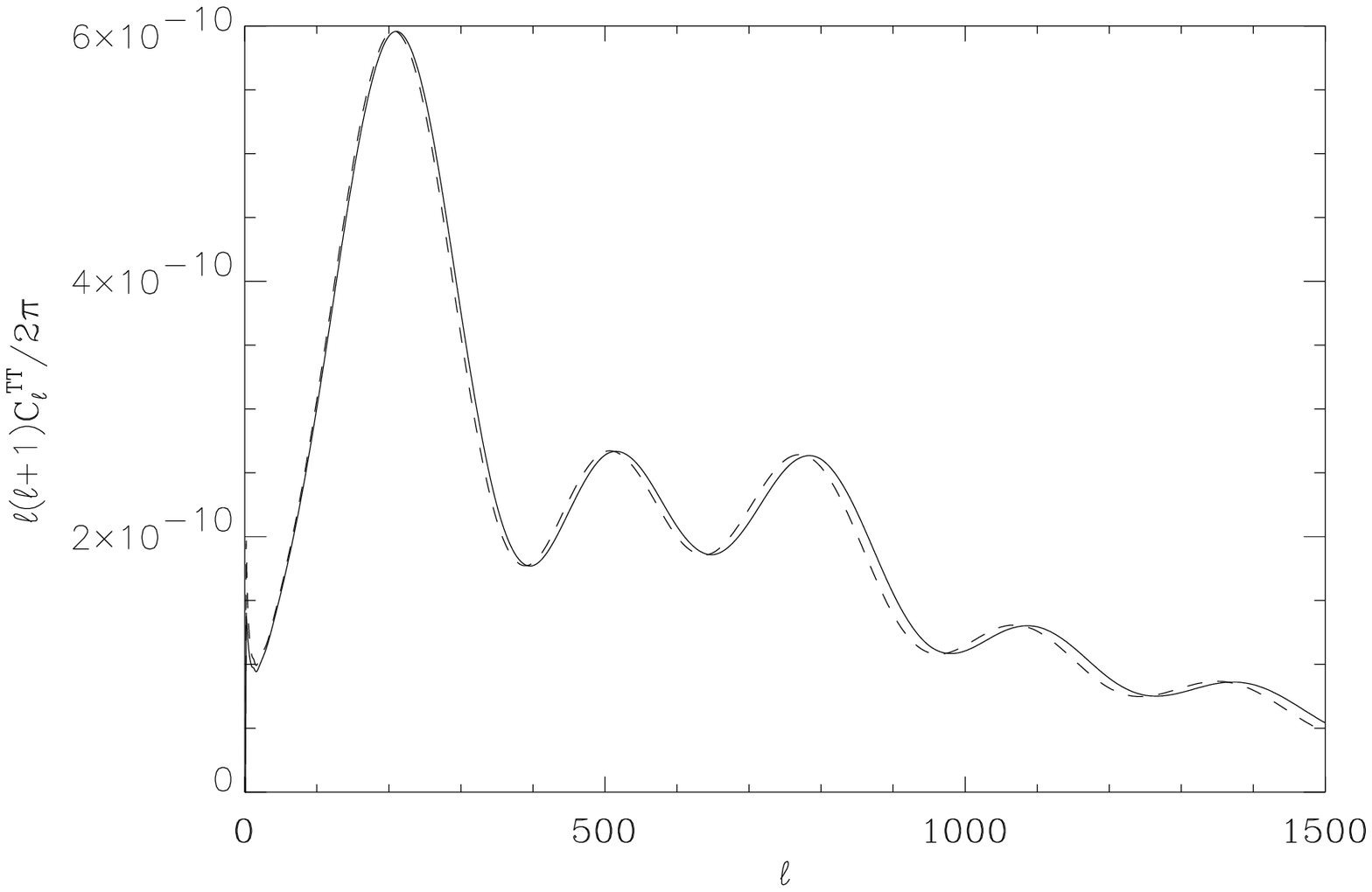}
\includegraphics[width=7cm,height=5cm]{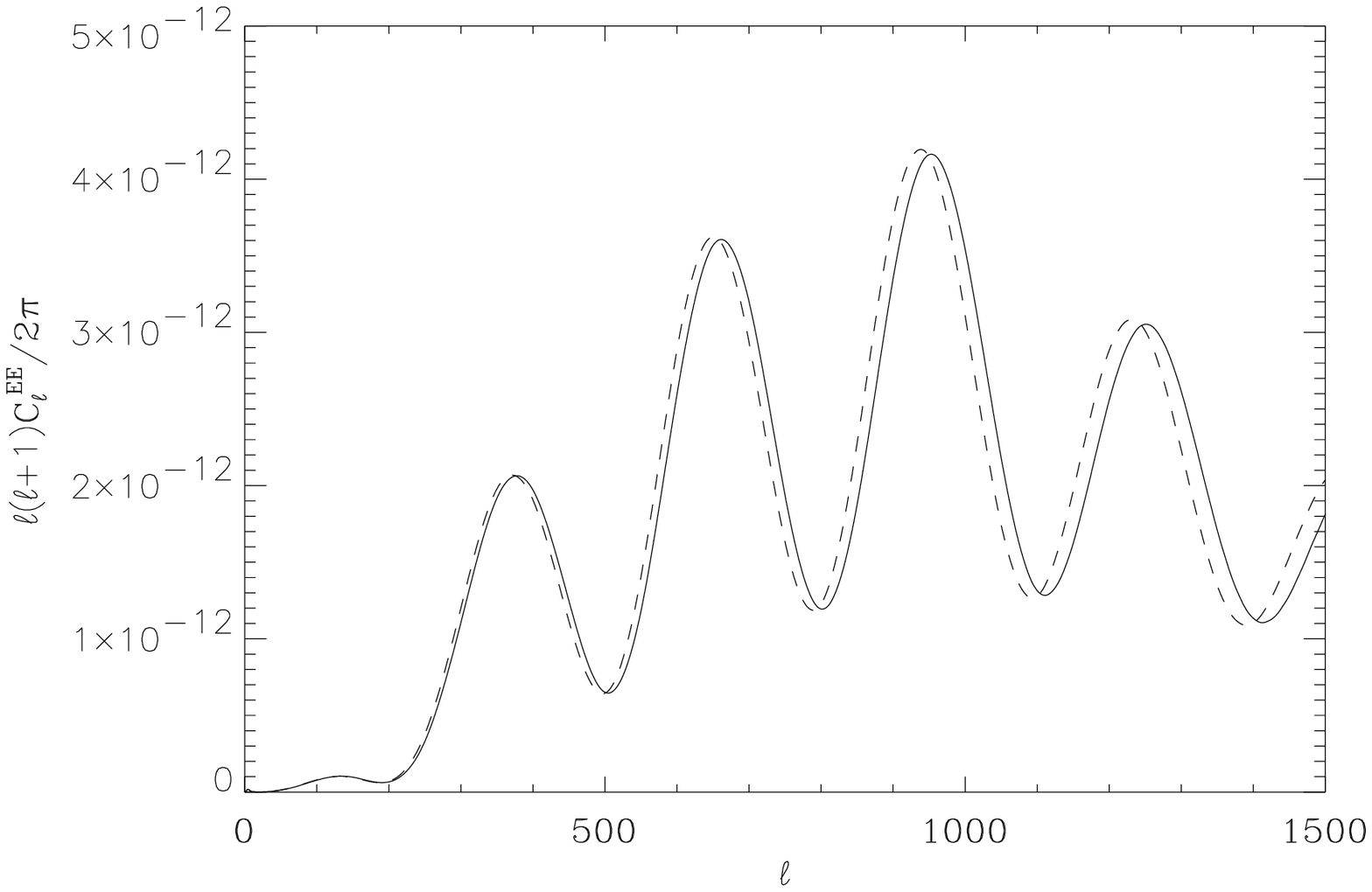}
\includegraphics[width=7cm,height=5cm]{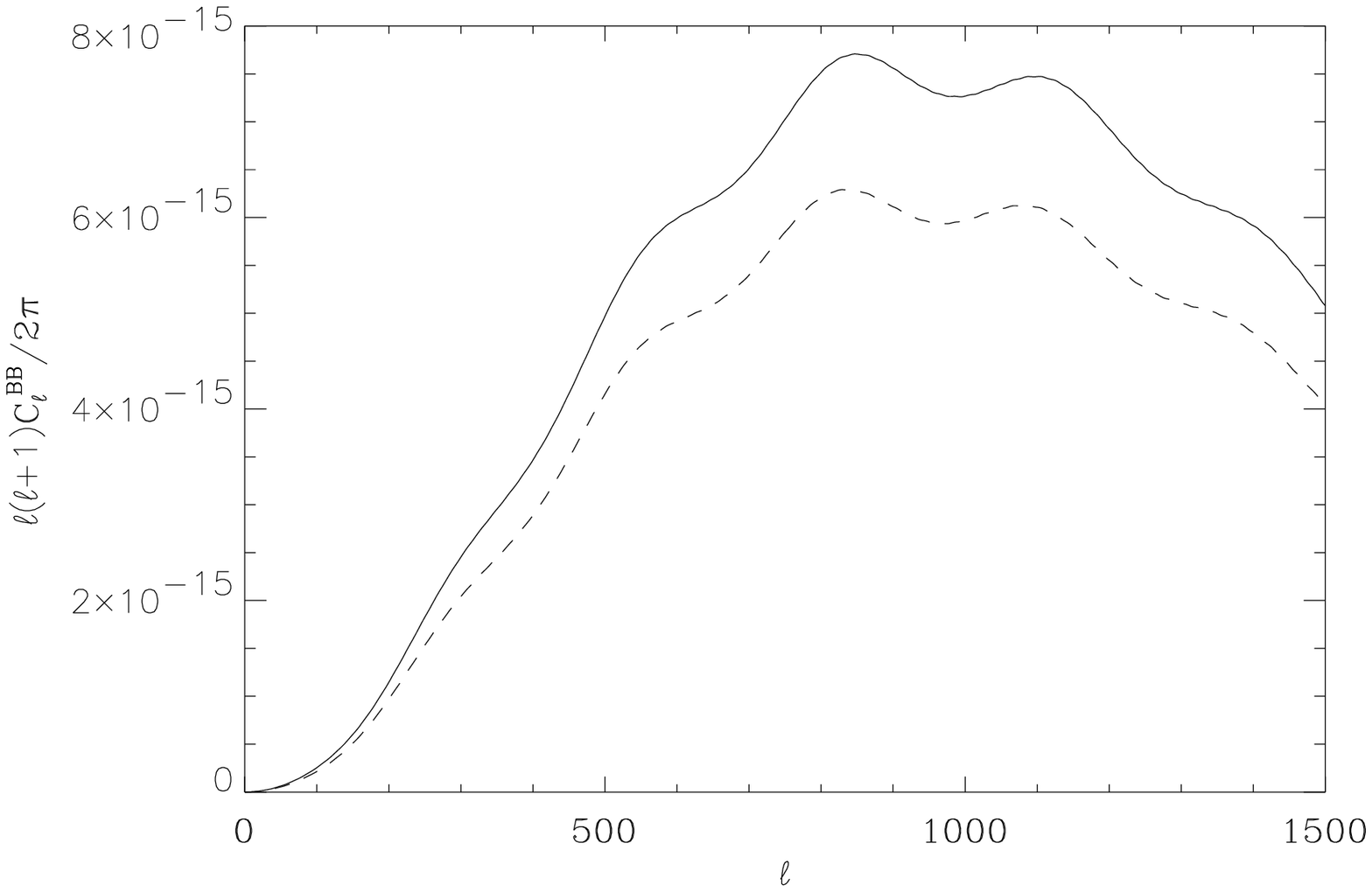}
\includegraphics[width=7cm,height=5cm]{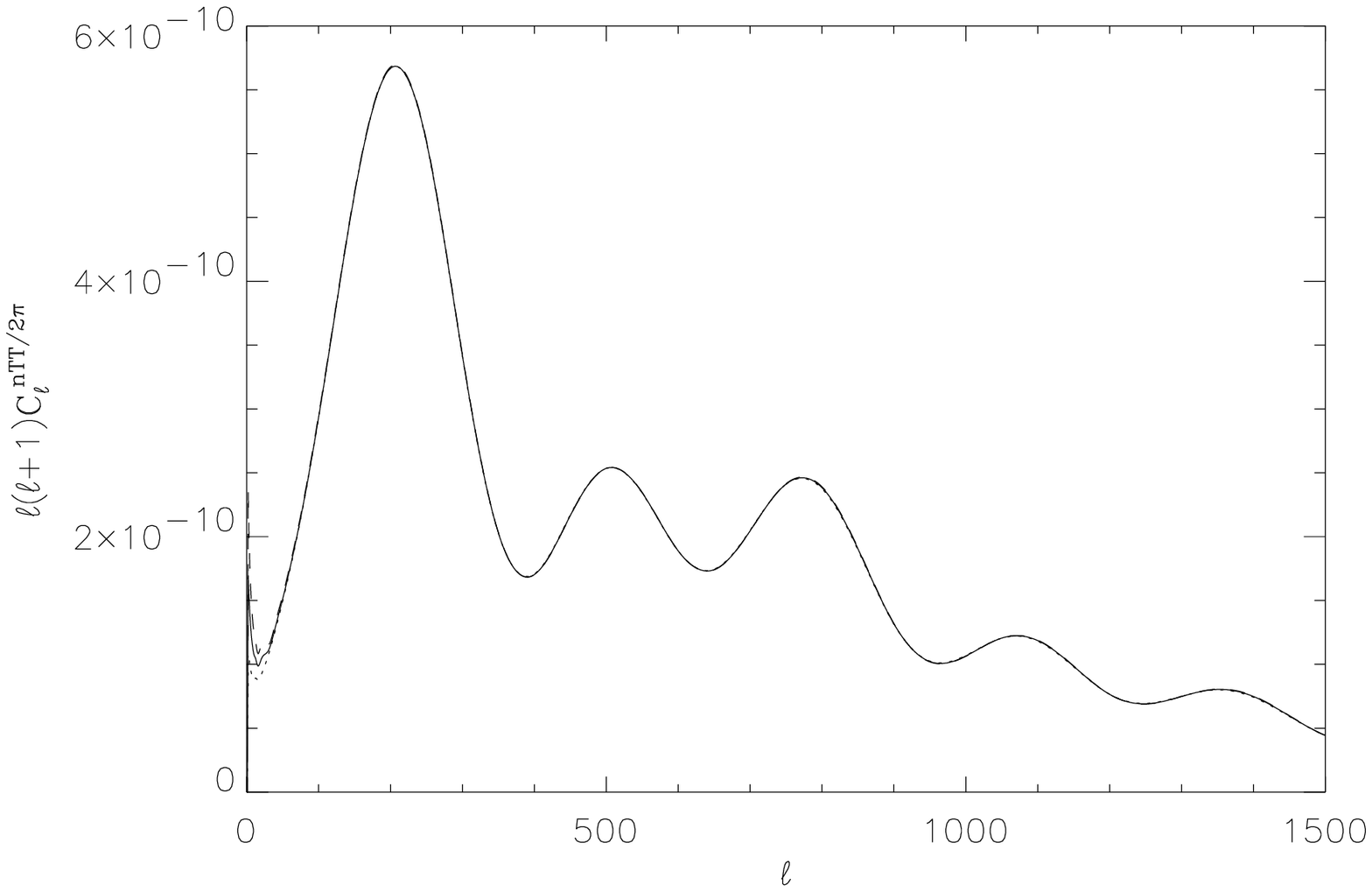}
\includegraphics[width=7cm,height=5cm]{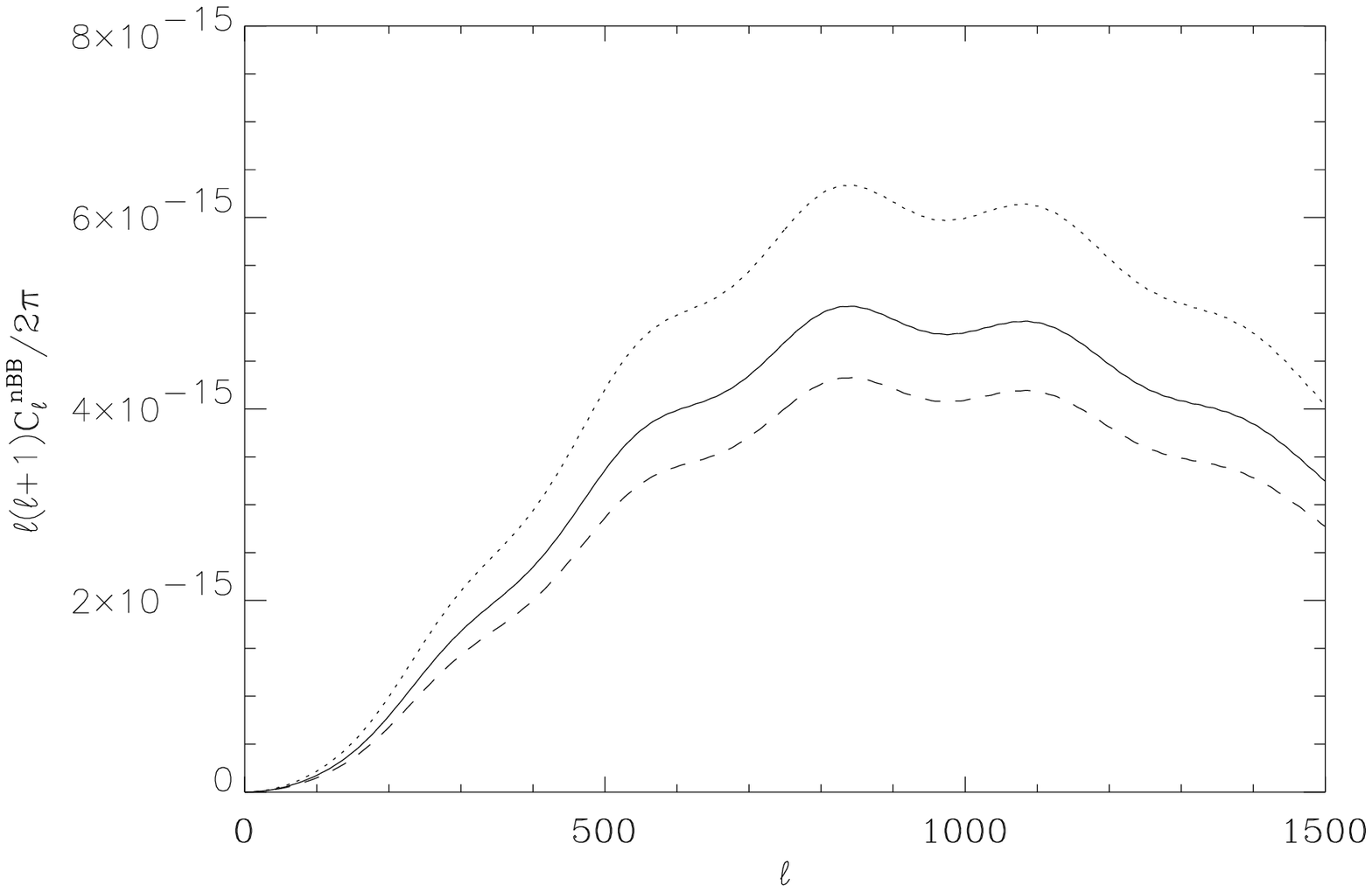}
\caption{Lensed CMB spectra in dark energy cosmologies with 
different equation of state behavior (top left): TT spectra ()
top right, EE spectra (middle left), BB spectra (middle right). 
Bottom: lensed TT (left) and BB (right) spectra for $\Lambda$CDM 
and two dark energy models featuring the same distance from last 
scattering.}
\label{SUGRARP}
\end{figure}

The generation of lensing BB modes is described in figure \ref{BB}. Even starting 
with a pure EE polarization at last scattering, a fraction of it becomes BB by lensing. 
Once again, this is due to the fact that CMB lensing is a second order effect for 
cosmological perturbations. Their independence at different wavelengths, 
valid at the linear level, no longer holds for lensing. 
The relevance of this effect on dark energy is also indicated in the figure, 
and is simply because the injection of $BB$ modes occurs at the onset of acceleration, and therefore should be sensitive to the dark energy abundance at that epoch. 
Another welcome feature is that the cosmological BB signal is expected to be dominated by lensing; any primordial power injected by gravitational waves peaks on the degree 
scale, and vanishes rapidly on smaller angles because of their relativistic behavior.\\
We now expose the relevant features of this argument, while a quantitative analysis 
can be found in \cite{acquaviva_baccigalupi_2005}. Let us consider two different 
dark energy models, featuring the same equation of state today, but markedly 
different in the past, as in figure \ref{SUGRARP}, top left panel; technically, 
those models correspond to a SUGRA (\cite{brax_martin_2000}, dashed line) and 
inverse power law quintessence potentials (\cite{ratra_peebles_1988}, solid line). They are suitable for our analysis, as they differ at high redshift. 
So we fix all cosmological parameters to be the same for the two 
models, including the primordial perturbation normalization, and we look at the 
lensed CMB angular power spectra in the top right and middle panels of figure \ref{SUGRARP}. 
While the TT and EE spectra undergo primarily a modest projection shift, 
the BB amplitude changes remarkably as a consequence of the modified 
value of $H$ at the epoch in which the lensing is injected, as a result 
of the different behavior in the two dark energy models. This is simply 
understood as follows: for a fixed primordial normalization, an higher 
value of $H$ results in an higher damping of perturbations because of 
cosmological friction, reducing the overall power of lensing. 
Indeed, the difference in amplitude of the BB modes traces the one in the value 
of $w$ at about redshift $1$; for instance, the height of the lensing peak shows 
a $30\%$ variation for the models in figure \ref{SUGRARP}, see the top left and 
middle right panel. \\
In order to demonstrate that this 
feature breaks the projection degeneracy mentioned above, in the two 
bottom panels of figure \ref{SUGRARP} we consider two different 
dark energy models with different values of $w_{0}$ and $w_{\infty}$ 
giving the same distance to last scattering (\ref{distance}) of 
a $\Lambda$CDM model, indicated as a dotted line. As it is evident, 
while the TT spectra are identical, the amplitude in BB changes remarkably. \\
As we stressed already, the quantitative assessment of how much 
this effect helps the determination of the redshift behavior from 
CMB lensing is ongoing \cite{acquaviva_baccigalupi_2005}. On the 
other hand, the good amplitude of the effect, the enhanced sensitivity 
to the dark energy abundance at the onset of acceleration, 
independently on the present, and ultimately the easy explanation 
in terms of the known cosmological dynamics makes this feature 
interesting. If the amplitude of the lensing BB modes will ever 
be measured, it will be straightforward to check if the 
level is the same as predicted in a $\Lambda$CDM or different, 
as models with $w>-1$ at the onset of acceleration predict. 

\section{Future CMB data and dark energy}
\label{fcmbdade}

\begin{figure}
\centering
\includegraphics[width=12cm,height=7cm]{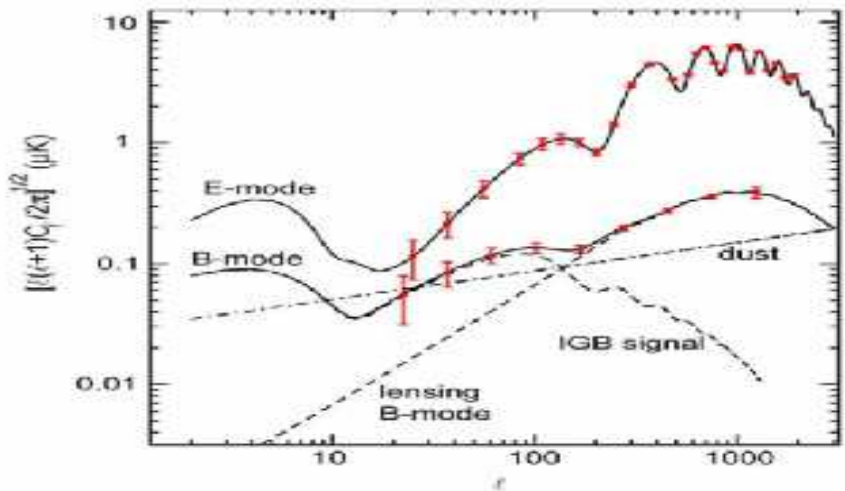}
\caption{The expected performance of the EBEx experiment for 
CMB polarization measurements \cite{oxley_etal_2004}.}
\label{EBEx}
\end{figure}

We conclude making some comments on the possibility to detect the 
lensing on the CMB in the near future. The attempts conducted 
so far on the data by the Wilkinson Microwave Anisotropy Probe 
(WMAP) were unsuccessful \cite{hirata_etal_2004}, while for 
the other experiments the lensing detection is hopeless due to 
instrumental limitations. \\
The forthcoming CMB probes are expected to have the instrumental 
capabilities to detect the lensing distortion in the CMB anisotropies, 
and in particular in the BB power in the polarization. \\
The Planck satellite (www.rssd.esa.int/Planck) has the appropriate 
sensitivity in order to detect the non-Gaussian power induced by lensing
on TT anisotropies; as any lensing observable, such non-Gaussian power is 
also injected at the onset of acceleration only. The potential relevance of 
this effect for constraining dark energy has also been addressed 
(see \cite{giovi_etal_2005} and references therein). \\
The CMB polarization on degree and sub-degree angular scales, and the 
BB modes in particular, are the main targets of the next sub-orbital CMB 
probes, either ground based and balloon borne, in a similar way as it was 
for the detection of the first peak in the CMB TT anisotropies, 
just a few years ago. There is no space to review all of them here, 
but a good list can be found at lambda.ngsc.gov where one can 
appreciate their number, and the consequent large theoretical and 
technological effort in this direction. As an example, in figure 
\ref{EBEx} we show the expected performance of the balloon borne 
E and B experiment (EBEx, see \cite{oxley_etal_2004}). 
The error bars are calculated based on the expected detector sensitivity, 
number of detectors, sky coverage and integration time. They do not include calibration or other systematic uncertainties. The total BB signal, 
indicated as a solid line as the EE one, is made by a primordial 
contribution of tensor modes from the Inflationary Gravitational 
Background (IGB), plus the lensing power. As it can 
be seen, in principle that instrument should measure the level of 
the lensing BB power with good accuracy. For sure, such performance 
has to be revised in terms of systematics, but the figure also 
highlights one of the main obstacles to the observation of the 
cosmological BB power in the CMB, which is represented by the 
diffuse foreground emission from our own Galaxy. At the frequencies of 
the EBEx experiment, between 150 and 450 GHz, the main contaminant is 
represented by the thermal dust emission, made by magnetized grains which 
get locally aligned by the Galactic magnetic field; its expectation at 150 
GHz is shown as a dashed line in the figure. While for CMB TT the sky 
is relatively free of foreground emission when the observation is 
targeted on a region far from the Galactic plane, for polarization, 
and BB modes in particular, the foreground contamination is potentially 
relevant on all sky regions, and at all frequencies, as the WMAP data 
recently confirmed \cite{page_etal_2006}. On the other hand, the statistical 
difference between background and foregrounds might be exploited 
in order to reduce the foreground contamination in the 
forthcoming CMB experiments, see \cite{stivoli_etal_2005} and references 
therein. 
 
In conclusion, we brought arguments here which show that the CMB has to be 
regarded as a probe of the differential behavior of the dark energy component, 
not only of its average as it has been so far. Its efficiency will be 
limited by instrumental systematics or foreground emission, but whether or 
not this will be worse than for other probes, e.g. supernovae or lensing of 
galaxies etc., will be clear only when the data will be actually taken. 

\begin{thebibliography}{99}
\bibitem{riess_etal_1998}
A.G.~Riess \etal, \emph{Observational Evidence from Supernovae for an Accelerating Universe and a Cosmological Constant}, \emph{\apj} {\bf 116} (1998) 1009[{\tt astro-ph/9805201}].
\bibitem{perlmutter_etal_1999}
S.~Perlmutter \etal, \emph{Measurements of Omega and Lambda from 42 High-Redshift Supernovae}, \emph{\apj} {\bf 517} (1999) 565 [{\tt astro-ph/9812133}].
\bibitem{spergel_etal_2006}
D.N.~Spergel \etal, \emph{Wilkinson Microwave Anisotropy Probe (WMAP) Three Year Results: Implications for Cosmology}, submitted to \emph{\apj} (2006) [{\tt astro-ph/0603449}].
\bibitem{ratra_peebles_1988}
B.~Ratra, P.J.E.~Peebles, \emph{Cosmological consequences of a rolling homogeneous scalar field}, \emph{\prd} {\bf 37} (1988) 3406.
\bibitem{wetterich_1988}
C.~Wetterich, \emph{Cosmologies with variable Newton's ``constant''}, \emph{\npb} {\bf 302} 645 (1988).
\bibitem{peebles_ratra_2003}
P.J.E.~Peebles, B.~Ratra, \emph{The cosmological constant and dark energy}, \emph{\rmp} {\bf 75} (2003) 559 [{\tt astro-ph/0207347}].
\bibitem{crittenden_pogosian_2006}
R.G.~Crittenden, L.~Pogosian, \emph{Investigating dark energy experiments with principal components}, (2005) [{\tt astro-ph/0510293}].
\bibitem{dick_etal_2006}
J.~Dick, L.~Knox, M.~Chu, \emph{Reduction of Cosmological Data for the Detection of Time-varying Dark Energy Density}, (2006) [{\tt astro-ph/0603247}].
\bibitem{chevallier_polarski_2001}
D.~Chevallier, D.~Polarski, \emph{Accelerating Universes with Scaling Dark Matter}, \emph{\ijmp} {\bf 10} (2001) 213 [{\tt gr-qc/0009008}].
\bibitem{linder_2003}
E.~V.~Linder, \emph{Exploring the Expansion History of the Universe}, \emph{\prl} {\bf 90} (2003) 091301 [{\tt astro-ph/0208512}].
\bibitem{seljak_etal_2005}
U.~Seljak \etal, \emph{Cosmological parameter analysis including SDSS Ly$\alpha$ forest and galaxy bias: Constraints on the primordial spectrum of fluctuations, neutrino mass, and dark energy}, \emph{\prd} {\bf 10} (2005) 103515 [{\tt astro-ph/0407372}].
\bibitem{doran_etal_2001}
M.~Doran, M.~Lilley, J.~Schwindt, C.~Wetterich, \emph{Quintessence and the Separation of Cosmic Microwave Background Peaks}, \emph{\apj} {\bf 559} (2001) 501 [{\tt  astro-ph/0012139}].
\bibitem{white_2005}
M.~White, \emph{Baryon oscillations}, \emph{\ap} {\bf 24} (2005) 334 [{\tt astro-ph/0507307}].
\bibitem{pogosian_etal_2005}
L.~Pogosian, P.S.~Corasaniti, C.~Stephan-Otto, R.~Crittenden, R.~Nicol, \emph{Tracking dark energy with the integrated Sachs-Wolfe effect: Short and long-term predictions}, \emph{\prd} {\bf 72} (2005) 103519 [{\tt astro-ph/0506396}].
\bibitem{hu_jain_2004}
W.~H, B.~Jain, \emph{Joint galaxy-lensing observables and the dark energy}, \emph{\prd} {\bf 70} (2004) 043009 [{\tt astro-ph/0312395}].
\bibitem{acquaviva_baccigalupi_2005}
V.~Acquaviva, C.~Baccigalupi, \emph{Dark energy records in lensed cosmic microwave background}, submitted to \emph{\prd} (2005) [{\tt astro-ph/0507644}].
\bibitem{bartelmann_schneider_2001}
M.~Bartelmann, P.Schneider, \emph{Weak gravitational lensing}, \emph{\pr} {\bf 340} (2001) 291 [{\tt astro-ph/9912508}].
\bibitem{lewis_challinor_2006}
A.~Lewis, A.~Challinor \emph{Weak gravitational lensing of the CMB}, \emph{\pr} {\bf 429} (2006) 1[{\tt astro-ph/0601594}].
\bibitem{brax_martin_2000}
P.~Brax, J.~Martin, \emph{Robustness of quintessence}, \emph{\prd} {\bf 62} (2000) 103502 [{\tt astro-ph/9912046}].
\bibitem{hirata_etal_2004}
C.M.~Hirata, N.~Padmanabhan, U.~Seljak, D.~Schlegel, J.~Brinkmann \emph{Cross-correlation of CMB with large-scale structure: Weak gravitational lensing}, \emph{\prd} {\bf 70} (2004) 103501 [{\tt astro-ph/0406004}].
\bibitem{giovi_etal_2005}
F.~Giovi, C.~Baccigalupi, F.~Perrotta, \emph{Cosmic microwave background constraints on dark energy dynamics: Analysis beyond the power spectrum}, \emph{\prd} {\bf 71} (2005) 103009 [{\tt 103009}].
\bibitem{oxley_etal_2004}
P.~Oxley \etal, \emph{The EBEX experiment}, \emph{Earth Observing Systems IX. Edited by Barnes, William L.; Butler, James J. Proceedings of the SPIE} {\bf 5543} (2004) 320 [{\tt astro-ph/0501111}].
\bibitem{page_etal_2006}
L.~Page \etal, \emph{Three Year Wilkinson Microwave Anisotropy Probe (WMAP) Observations: Polarization Analysis}, submitted to \emph{\apj} (2006) [{\tt astro-ph/0603450}].
\bibitem{stivoli_etal_2005}
F.~Stivoli, C.~Baccigalupi, D.~Maino, R.~Stompor, \emph{Separating cosmological B modes from foregrounds in cosmic microwave background polarization observations}, submitted to \emph{\mnras} (2005) [{\tt astro-ph/0505381}].
\end{thebibliography}
\end{document}